\documentclass[journal]{IEEEtran}
\makeatletter
\def\endthebibliography{%
 \def\@noitemerr{\@latex@warning{Empty `thebibliography' environment}}%
 \endlist
}
\makeatother
\usepackage{amsfonts}
\usepackage{mathrsfs}
\usepackage{amsmath, amssymb, amsthm, latexsym, amsfonts, epsfig, color,graphicx}

\usepackage{subfigure}
\usepackage{epstopdf}
\usepackage{amsmath}
\usepackage{cases,bm}
\usepackage{graphics}
\usepackage{makecell}

\usepackage[numbers,sort&compress]{natbib}

\usepackage[colorlinks, linkcolor=blue, anchorcolor=blue, citecolor=blue]{hyperref}

\usepackage{lipsum}

\usepackage{algorithm}
\usepackage{algorithmic}

\newcommand{\eproof}{\hfill\rule{2mm}{2mm}}

\newcommand{\proofnow}{\newcommand{\eproof}{\hfill\rule{2mm}{2mm}}

\newcommand{\proofnow}{\noindent{\bf Proof: }}{\bf Proof: }}

\newtheorem{Theorem}{Theorem}

\newtheorem{Corollary}[Theorem]{Corollary}

\newtheorem{Remark}{Remark}

\numberwithin{equation}{section}
\allowdisplaybreaks

 \makeatletter\def\theequation{\arabic{section}.\arabic{equation}}\makeatother

\begin{document}

%
\title{Adaptive support driven Bayesian reweighted algorithm for sparse signal recovery}
\author{Junlin Li, Wei Zhou and Cheng Cheng$^{\dagger}$
\thanks{$^\dagger$Corresponding author. J. Li, W. Zhou and C. Cheng are with the School of Artificial Intelligence and Automation, Huazhong University of Science and Technology, Wuhan, 430074, China.}}
\maketitle

\begin{abstract}
Sparse learning has been widely studied to capture critical information from enormous data sources in the filed of system identification. Often, it is essential to understand internal working mechanisms of unknown systems (e.g. biological networks) in addition to input-output relationships. For this purpose, various feature selection techniques have been developed. For example, sparse Bayesian learning (SBL) was proposed to learn major features from a dictionary of basis functions, which makes identified models interpretable. Reweighted $\ell_1$-regularization algorithms are often applied in SBL to solve optimization problems. However, they are expensive in both computation and memory aspects, thus not suitable for large-scale problems. This paper proposes an adaptive support driven Bayesian reweighted (ASDBR) algorithm for sparse signal recovery. A restart strategy based on shrinkage-thresholding is developed to conduct adaptive support estimate, which can effectively reduce computation burden and memory demands.
 Moreover, ASDBR accurately extracts major features and excludes redundant information from large datasets. Numerical experiments demonstrate the proposed algorithm outperforms state-of-the-art methods.
\end{abstract}

\begin{IEEEkeywords}
Sparse Bayesian Learning, noncovex optimization, sparse signal recovery.
\end{IEEEkeywords}

\section{Introduction}
Sparse signal recovery has been widely studied due to its potential application in the area of signal processing, system identification and machine learning \cite{donoho2006compressed,zhang2011sparse,eldar2012compressed}. The canonical form of this problem is given by
\begin{equation}\label{eq0}
y=\Phi \theta+w,
\end{equation}
where $\Phi\in \mathbb{R}^{m\times n}$ is the measurement or design matrix, $y\in \mathbb{R}^{m}$ is the observation vector, $\theta\in \mathbb{R}^{n}$ is the unknown coefficient vector to be learned, $w\in \mathbb{R}^{m}$ is zero-mean additive observation noise. In the sparse recovery problem, we seek a parameter vector $\theta$ whose entries are predominantly zero to accurately approximate $y$. This is equivalent to representing $y$ with a minimal number of basis functions.

A natural optimization-theoretic formulation of \eqref{eq0} is via $\ell_0$-minimization, where $\ell_0$-norm of a vector represents the number of nonzero entries. However, since $\ell_0$-norm is nonconvex, it is intractable to solve $\ell_0$-minimization problem especially when $n$ is large \cite{nat1995}.
 To alleviate such an issue, an effective way  is to replace the troublesome $\ell_0$-norm by computationally tractable approximations or relaxations. Specially, the $\ell_1$-norm is frequently used as the optimal convex surrogate of $\ell_0$-norm over the past decades, which leads to the following optimization problem
\begin{equation}\label{p1}
 \min \limits _{\theta\in \mathbb{R}^n}\frac {1}{2}\|y-\Phi\theta\|^2_2+\lambda\|\theta\|_1,
\end{equation}
which is known as Lasso \cite{tib1996}.
It is worthy noting that Lasso estimate could be interpreted as maximum a posteriori estimate (MAP) in the linear model \eqref{eq0} with a Laplace prior on the coefficient vector \cite{park2008bayesian,rish2014}.

Because of the $\ell_1$-geometry, Lasso is often employed to estimate the coefficients with a sparse structure.
In practice, the problem can be solved using quadratic programming approach \cite{tur2005}, coordinate wise optimization \cite{fri2007}, or gradient projection method \cite{fig2007}. Moreover, some theoretical results were established to show that coefficients estimated by Lasso are consistent \cite{wan2009,zhao2006}.
However, the $\ell_1$-norm is not the best approximation of $\ell_0$-norm. In \cite{fan2001}, Fan and Li proposed some arguments against the Lasso because the
$\ell_1$-norm penalty associated with the Lasso produces biased estimates for large coefficients. Thereby, other penalty functions that lead to sparse and unbiased models are utilized to replace the $\ell_1$-norm. To this end, they advocate that the penalty functions should be singular at the origin and their derivatives should vanish for large values in order to achieve sparsity.
In particular, such a nonconvex penalty can be obtained in SBL framework, which has been verified to outperform $\ell_1$-norm for sparse approximation or promoting sparsity \cite{fau2002,pan2015sparse,yuan2019machine}.  

In SBL, a sparsity-inducing prior with a set of hyperparameters is imposed on the coefficient vector to achieve a sparse model.
The unknown hyperparameters are estimated by evidence maximization.
As a pioneer work, an SBL algorithm based a Gaussian-inverse Gamma model was developed via expectation maximization~\cite{tip2001}.
In~\cite{wip2004}, theoretical analysis was provided to show that SBL can produce sparse solutions.
It was demonstrated in~\cite{ji2008bayesian} that the sparse signal recovery problem can be solved more effectively in the sparse Bayesian framework.
Subsequently, a Laplace prior model was employed to induce a sparse model via an SBL procedure~\cite{babacan2009bayesian}.
Based on a power exponential scale mixture prior, two types of Bayesian methods were presented for sparse signal recovery~\cite{giri2016type}, which establishes a connection between $\ell_1$-norm minimization methods and SBL approaches.
Along this research line, the SBL optimization problem was solved by a reweighted $\ell_1$-minimization algorithm \cite{wipf2008new,yuan2019data}, where the coefficient vector is updated based on the previous estimates of the coefficient vector and hyperparameter vector.
However, this algorithm is expensive in computation and memory, especially for large-scale problems.


In this paper, we propose an adaptive support driven Bayesian reweighted (ASDBR) algorithm. In ASDBR, the original $\ell_1$-minimization problem is replaced by a sequence of reweighted $\ell_1$-minimization subproblems with iteratively updated weights applied to the adaptive support estimate. Therein, the reweighted $\ell_1$-minimization subproblems can be solved by shrinkage-thresholding algorithm (ISTA) previously proposed in \cite{daubechies2004iterative,beck2009fast}. ASDBR has two main parts: adaptive support estimate and iteratively updated weights. Using the proposed threshold strategy, the support estimate can be obtained in each outer iteration of ASDBR. The updated weights depend on the previous estimates of coefficient vectors and hyperparameter vectors in the support estimate. Monte Carlo simulations show that ASDBR can increase computational speed with low memory consumption compared with state-of-the-art methods.

The structure of this paper is organized as follows. In section \ref{se2}, some preliminaries are briefly reviewed. In section \ref{method}, the ASDBR algorithm is developed. In section \ref{se4}, numerical experiments are implemented to demonstrate the effectiveness of the proposed algorithm. Finally, conclusions are drawn in Section \ref{se5}.

\section{Preliminaries}\label{se2}

\subsection{Notations }
For a vector $x\in \mathbb{R}^n$, $x_i$ stands for the \emph{i}th entry of $x$, $\mathrm{diag}(x)$ represents a $n\times n$ square diagonal matrix with the elements of vector $x$ on the main diagonal.
For a matrix $A\in \mathbb{R}^{m\times n}$, $A_i$ denotes the \emph{i}th column of $A$. For a set $S\subseteq \{1,2,\cdots,n\}$, $|S|$ denotes the cardinality of $S$. We use $A_S$ to denote the $m\times|S|$ submatrix of $A$ containing the columns indexed by $S$. Similarly, $x_S$ denotes the subvector of $x$ containing the entries indexed by $S$. $(\cdot)^T$ represents the transpose. $\|x\|_2$ and $\|x\|_1$ represent the $\ell_2$ and $\ell_1$ norms of vector $x$, respectively.

\subsection{Iterative shrinkage-thresholding}
Iterative shrinkage-thresholding algorithms (ISTA) can be viewed as a special proximal forward backward iterative scheme introduced in \cite{bruck1977weak} and \cite{passty1979ergodic}.  Assume that $\left\{\tau^{(k)}\right\}_{k \in \mathbb{N}}$ is a positive real number sequence satisfying $\inf _{k \in \mathbb{N}} \tau^{(k)}>0$ and $\sup _{k \in \mathbb{N}} \tau^{(k)}<\|\Phi\|_{2}^{-2}$, $\left\{u^{(k)}\right\}_{k \in \mathbb{N}}$ is a sequence
in $\mathbb{R}^{n}$. Then, the general step of ISTA is
\begin{equation}\label{ist1}
  x^{(k+1)}=\eta\left(x^{(k)}+\tau^{(k)} \Phi^{\top}(y-\Phi x^{(k)})+\tau^{(k)}u^{(k)}, \lambda \tau^{(k)}\right),
\end{equation}
where $\eta(x,\nu)$ is a soft-thresholding operator defined by
\[
\eta(x,\nu)=\left\{\begin{array}{ll}
{\operatorname{sgn}(x)(|x|-\nu),} & {\text { if }|x|>\nu} \\
{0,} & {\text { otherwise }}
\end{array}\right.
\]
which is applied component-wise.
The original ISTA, previously proposed in \cite{daubechies2004iterative}, has the form \eqref{ist1} with $u^{(k)}=0$ and $\tau^{(k)}=\tau<\|\Phi\|_{2}^{-2}$ for all $k \in \mathbb{N},$ which can be guaranteed to converge to a solution of \eqref{p1} under some assumptions. The advantages of this algorithm lies in its simplicity for high-dimensional problems.
\cite{bredies2007iterative} showed that this algorithm converges with linear rate under some assumptions.
Moreover, some techniques can be utilized to improve this algorithm. In \cite{maleki2010optimally}, the authors proposed a parameter tuning scheme in terms of phase transitions, i.e., maximize the number of nonzeros at which the algorithm can successfully operate. However, this scheme is expensive and has no theoretical guarantee. Other update schemes for the next iteration not only depend on the current estimate, but also previously computed estimates. For example, Fast Iterative Shrinkage-Thresholding Algorithm (FISTA) \cite{beck2009fast} obtained by \eqref{ist1} choosing $\tau^{(k)}=\tau<\|\Phi\|_{2}^{-2}$ and
\[
\begin{array}{l}
{u^{(k)}=\frac{t^{(k-1)}-1}{t^{(k)}}\left(\tau^{-1} I-\Phi^{\top} \Phi\right)\left(x^{(k)}-x^{(k-1)}\right)} \\
{t^{(0)}=1, \quad t^{(k+1)}=\frac{1+\sqrt{1+4\left(t^{(k)}\right)^{2}}}{2}}.
\end{array}
\]

Specially, for the reweighted $\ell_1$-minimization problem
\begin{equation}\label{ist2}
 \min \limits _{\theta\in \mathbb{R}^n}\frac {1}{2}\|y-\Phi\theta\|^2_2+\lambda\|W\theta\|_1,
\end{equation}
where $\mathbb{R}^{n\times n}\ni W=\mathrm{diag}(w)\succ 0$ is a diagonal matrix. It can be solved by the the iterative schemes:  $$x^{(k+1)}=\eta\left(x^{(k)}+\tau^{(k)} \Phi^{\top}(y-\Phi x^{(k)})+\tau^{(k)}u^{(k)}, \lambda \tau^{(k)}w\right).$$


\section{Method for sparse recovery}\label{method}
\subsection{Problem formulation}
 For the linear model \eqref{eq0}, assume that the noise vector $w$ follows Gaussian distribution $\mathcal{N}(0,\lambda I_m)$. Then the likelihood of the target given the $\theta$ is
\begin{equation}\label{eq1}
  p(y|\theta)=(2\pi\lambda)^{-\frac {m}{2}}\exp\left(-\frac {1}{2\lambda}\|y-\Phi\theta\|^2_2\right).
\end{equation}
Under the Bayesian paradigm, unknown parameters of model \eqref{eq0} are treated as random variables. Specially, the parametric form of the coefficient prior is given by
\begin{equation}\label{eq2}
 p(\theta;\gamma)=\prod_{i=1}^{n}(2\pi\gamma_i)^{-\frac {1}{2}}\exp\left( -\frac {\theta_i^2}{2\gamma_i}\right),
\end{equation}
where $\gamma:=[\gamma_1,\gamma_2,\cdots,\gamma_n]^T$ is a vector of $n$ hyperparameters determining the variance of each coefficient. Then, the objective is to optimize variables $\theta$ given measurements.
 \subsection{Bayesian inference}
 These hyperparameters can be estimated by a type-$\Pi$ maximum likelihood method \cite{wipf2008new,pan2015sparse}, i.e., marginalizing over the coefficients and then performing maximum likelihood optimization. The marginalized probability distribution function is given by
\begin{align}\notag
  p(y;\gamma)&=\displaystyle{\int p(y|\theta)p(\theta;\gamma)d\theta} \\ \label{eq3}
  &=(2\pi)^{-\frac{m}{2}}|\Sigma_y|^{-\frac {1}{2}}\exp\left(-\frac {1}{2}y^T\Sigma_y^{-1}y\right),
\end{align}
where $\Sigma_y:=\lambda I_m+\Phi\Gamma\Phi^T$ and $\Gamma:=\mathrm{diag}(\gamma_1,\cdots,\gamma_n)$. Then $\gamma$ can be estimated via maximum likelihood. This is equivalent to minimizing $-\log p(y;\gamma)$, giving the Bayesian cost function
\begin{equation}\label{eq6}
 \mathcal{ L}_\gamma(\gamma)=\log|\Sigma_y|+y^T\Sigma_y^{-1}y,
\end{equation}
that is,
\begin{equation}\label{eq7}
\mathcal{L}_\gamma(\gamma)=\log|\lambda I_m+\Phi\Gamma\Phi^T|+y^T(\lambda I_m+\Phi\Gamma\Phi^T)^{-1}y.
\end{equation}
Note that
\begin{align*}
   y^T(\lambda I_m+\Phi\Gamma\Phi^T)^{-1}y =& \frac{1}{\lambda}y^Ty-\frac{1}{\lambda^2}y^T\Phi\Sigma_\theta\Phi^Ty\\
      =& \frac{1}{\lambda}\|y-\Phi\mu_\theta\|^2_2+\mu_\theta^T\Gamma^{-1}\mu_\theta\\
            =& \min \limits_{\theta}\Big{\{} \frac{1}{\lambda}\|y-\Phi\theta\|^2_2+\theta^T\Gamma^{-1}\theta \Big{\}}.
\end{align*}
For fixed values of the hyperparameters, the posterior density of the coefficients is Gaussian, i.e.,
\begin{equation}\label{eq4}
  p(\theta|y;\gamma)=\mathcal{N}(\mu_\theta,\Sigma_\theta)
\end{equation}
with $\mu_\theta=\lambda^{-1}\Sigma_\theta\Phi^Ty$ and $\Sigma_\theta=(\lambda^{-1}\Phi^T\Phi+\Gamma^{-1})^{-1}$. Thereby, once we obtain the estimate $\widehat{\gamma}$ for $\gamma$, we have the estimate for $\theta$:
\begin{equation}\label{eq5}
 \widehat{\theta}=\widehat{\mu}_\theta=(\Phi^T\Phi+\lambda \widehat{\Gamma}^{-1})^{-1}\Phi^Ty,
\end{equation}
where $\widehat{\Gamma}=\mathrm{diag}(\widehat{\gamma}_1,\widehat{\gamma}_2,\cdots,\widehat{\gamma}_n)$. Thus, the estimate for $\theta$ can be obtained by solving the optimization problem
\begin{equation}\label{eq7.1}
  \min \limits_{\theta} \|y-\Phi\theta\|^2_2+\lambda \psi(\theta),
\end{equation}
where $\psi(\theta):=\min \limits_{\gamma\succeq 0}\{ \theta^T\Gamma^{-1}\theta+ \log|\lambda I_n+\Phi\Gamma\Phi^T| \} $. Note that this problem does not have a closed-form solution. Moreover, \eqref{eq7.1} is equivalent to the following optimization problem
 \begin{equation}\label{eq7.22}
  \min \limits_{\theta,\gamma\succeq 0} \|y-\Phi\theta\|^2_2+\theta^T\Gamma^{-1}\theta+ \log|\lambda I_n+\Phi\Gamma\Phi^T|.
\end{equation}
This problem can be solved by the concave-convex procedure (CCP) as follows,
\begin{equation}\label{eq16.1}
  \theta^{(k+1)}\in \arg \min \limits_{\theta} \Big{\{}   \frac {1}{2}\|y-\Phi\theta\|^2_2+\lambda \|W^{(k)}\theta\|_1\Big{\}},   
\end{equation}
where $W^{(k)}:=\mathrm{diag}(\sqrt{c^{(k)}_1},\cdots,\sqrt{c^{(k)}_n})$, and
 \begin{equation*}
   c^{(k)}:=\nabla_{\gamma}\log|\lambda I_m+\Phi\Gamma^{(k)}\Phi^T|,
 \end{equation*}
\begin{equation*}
\gamma^{(k+1)}_j=\frac {|\theta_j^{(k+1)}|} {\sqrt{c_j^{(k)}}},\ j=1,2,\cdots,n.
\end{equation*}
 Moreover, $c^{(k)}$ can be calculated with
\begin{align}\notag
  c^{(k)}&=\nabla_\gamma \log|\lambda I_m+\Phi\Gamma^{(k)}\Phi^T|\\ \label{eq13}
           &=\mathrm{diag}[\Phi^T(\lambda I_m+\Phi\Gamma^{(k)}\Phi^T)^{-1}\Phi].
\end{align}

\subsection{Sparse recovery based on Bayesian reweighted algorithm}\label{se3}

As mentioned above, \eqref{eq16.1} can be regarded as an reweighted $\ell_1$-regularization problem with regularization parameter $\lambda$. In general, \eqref{eq16.1} is computed by using the third party solver, e.g., CVX.
In addition, using such program for sparse recovery will encounter two challenges.
\begin{itemize}
  \item [(1)] Based on \eqref{eq13}, matrix inversion is required to obtain the weights at each iteration, which leads to large computational complexity.
  \item [(2)] The memory consumption depends on the size of $\Phi^T\Phi$ (See \eqref{eq13}), which is costly for the large-scale measurement matrix $\Phi$.
\end{itemize}
To alleviate these challenges, we propose a novel algorithm called ASDBR that combines two strategies. The first one is the adaptive support ($AS$). The second one is to use ISTA (or its variant) to solve the reweighted $\ell_1$-minimization problem. The main steps of the proposed algorithm are summarized in Algorithm \ref{alg1}.
\begin{algorithm}
  \caption{ASDBR algorithm}
  \label{alg1}
\textbf{Input:}
      $\Phi\in \mathbb{R}^{ m\times n}$: design matrix;
      $y\in \mathbb{R}^{m}$: observation vector;
      $\tau$: tolerance; $k_{\rm  inner}$: maximum iteration number for ISTA; $k_{\rm  outer}$: maximum iteration number for ASDBR. \\
      \textbf{Unweighted $\ell_1$-minimization:}\\
       Compute the $\ell_1$-minimization problem \eqref{eq16.1} with $W^{(0)}=I_n$ using ISTA algorithm with $k_{\rm  inner}$ iterations to obtain an estimate $\theta^{(1)}$ of $\theta$.\\
       \textbf{Reweighted $\ell_1$-minimization:}
       \begin{itemize}
        \item[(1)] Initialize with $\Phi^{(1)}=\Phi$, $W^{(0)}_{\rm new}=I$, $S^{(0)}_0=\{1,\cdots,n\}$, $V^{(0)}=[1,2,\cdots,n]^T$, $S^{(0)}=\{1,\cdots,n\}$, $\widehat{\theta}=[0,\cdots,0]^T\in \mathbb{R}^n$, $k=1$.
        \item[(2)] In order to achieve the support set estimate for ASDBR, set $\theta^{(k)}_i=0$ when $|\theta^{(k)}|<0.01\times \max\{|\theta^{(k)}|\}$.
         \item[(3)] 
         Define the indices set $I^{(k)}:=\{i:\theta^{(k)}_i\neq0\}$, then we obtain the support set estimation $S^{(k)}:=\{V^{(k)}_i:i=1,\cdots,|V^{(k)}|\}$ with $V^{(k)}:=V^{(k-1)}_{I^{(k)}}$.
      \item[(4)] Generate the weighted matrix $W^{(k)}_{\rm new}$ by using $W^{(k)}_{\rm new}:=\mathrm{diag}(w^{(k)}_{I^{(k)}})$, $W^{(k)}:=\mathrm{diag}(w^{(k)})$,
             $w^{(k)}=\{(\Phi^{(k)})^T[\lambda I_{I^{(k-1)}}+\Phi^{(k)}\Gamma^{(k)}(\Phi^{(k)})^T]^{-1}\Phi^{(k)}\}^{1/2}$, $\Gamma^{(k)}=\mathrm{diag}(|\theta^{(k)}|)[(W^{(k-1)}_{\rm new}]^{-1}$.%
         \item[(5)] By removing columns of $\Phi^{(k)}$ corresponding to zero entries in $\theta^{(k)}_i$, $\Phi^{(k)}$ is compressed into $\Phi^{(k+1)}$. Then, the reweighed $\ell_1$-minimization problem turns into
         \[
        \theta^{(k+1)}\in \arg \min \limits_{\theta\in \mathbb{R}^{|I^{(k)}|}}\frac {1}{2}\|y-\Phi^{(k+1)}\theta\|^2_2+\lambda \|W^{(k)}_{new}\theta\|_1.
         \]
          \item[(6)] Solve the optimization problem by using ISTA algorithm with $k_{\rm inner}$ iterations and initial values $\theta^{(k+1)}=\theta^{(k)}_{I^{(k)}}$.
      \item[(7)] $k=k+1$. If $k>k_{\rm  outer}$ or $|S^{(k)}|=|S^{(k-1)}|$, quit the iteration.
       \end{itemize}
       \textbf{Output:}
       $\widehat{\theta}\in \mathbb{R}^n$ with $\widehat{\theta}_{S^{(k)}}=\theta^{(k+1)}$.
\end{algorithm}

In Algorithm \ref{alg1}, the solution $\theta^{(1)}$ of the unweighted $\ell_1$-minimization based on ISTA is used as the initial values of the reweighted $\ell_1$-minimization (RL1), and its support set is regard as the first support set estimate. Here, $k_{\rm inner}$ is used as the maximum inner iteration of ISTA.

In the step 2 and step 3 of RL1, the support set estimate is obtained by thresholding the absolute vector $|\theta^{(k)}|$ for avoiding to delete the correct support set existing in the indices corresponding to the negative values. The threshold strategy allows us to neglect some very small nonzero entries compared to those of entries. In general, the threshold is set to 0.01 in this work. Consider that the columns of the dictionary matrix $\Phi$ will be reduced at every iteration. To record the indices for the remaining columns, we define the indices set $I^{(k)}$ and the support set estimate vector $V^{(k)}$ at $k$th iteration. Moreover, the set of all entries of $V^{(k)}$ is called the support set estimate at $k$th iteration, and is denoted as $S^{(k)}$, i.e., $S^{(k)}:=\{V^{(k)}_i:i=1,\cdots,|V^{(k)}|\}$.
Note that, at the $k$th iteration, $V^{(k)}\in \mathbb{R}^{|I^{(k)}|}$ is obtained by mapping $V^{(k-1)}$ onto $I^{(k)}$, i.e., $V^{(k)}:=V^{(k-1)}_{I^{(k)}}$. In the step 4 of RL1, the sub-weighted matrix $W^{(k)}$ is constructed by the estimate $\theta^{(k)}$ and the dictionary matrix $\Phi^{(k)}$ at $k$th iteration. Here, we only consider the weighted matrix that its diagonal entries is the entries of $\theta^{(k)}_{I^{(k)}}$, denoted by $W^{(k)}_{\rm new}$, and then $W^{(k)}_{\rm new}=\mathrm{diag}(\theta^{(k)}_{I^{(k)}})$.
In the step 5 of RL1, we delete the columns of $\Phi^{(k)}$ whose column indices are not in $I^{(k)}$, and hence $\Phi^{(k)}$ is compressed into $\Phi^{(k+1)}$. Thereby, we solve the reweighted $\ell_1$-minimization problem with dictionary matrix $\Phi^{(k+1)}$ and weighted matrix $W^{(k)}_{\rm new}$. Note that, this optimization problem is established in $\mathbb{R}^{|I^{(k)}|}$.
In step 6 of RL1, the reweighted $\ell_1$-minimization problem is solved using ISTA with initial values $\theta^{(k+1)}_0=\theta^{(k)}_{I^{(k)}}\in\mathbb{R}^{|I^{(k)}|}$.
The step 7 of RL1 is to determine whether to terminate the proposed algorithm. Finally, the output vector $\widehat{\theta}$ should is $n$-dimensional vector, whose entries corresponding to the indices in support set estimate $S^{(k)}$ equal to the entries in $\theta^{(k+1)}$ and the rest of the components are zero. In fact, in ASDBR, the columns of the dictionary matrix $\Phi$ are pruned as the number of iterations increases. Note that, the threshold strategy is utilized to delete the term with small coefficients. And, $\theta^{(k)}$ generated at $k$th iteration is a $|I^{(k)}|$-dimensional vector, rather than a $n$-dimensional vector.

Consider that the cardinality of support set estimate $S^{(k)}$ decreases as the number of iterations $k$ increases. Since $0\leq |S^{(k)}|\leq n$, $k=1,\cdots$, the sequence $\{|S^{(k)}|\}$ will eventually stabilize at an integer. Here, an appropriate stopping criteria for ASDBR algorithm is that there exists an integer $k$ such that $|S^{(k)}|=|S^{(k-1)}|$. Then, we have the following result.

\begin{Theorem}\label{th1}
Let $\{S^{(k)}\}$ be a support set estimate sequence generated from ASDBR algorithm. Then, ASDBR algorithm terminates in at most $n+1$
iterations.
\end{Theorem}
\noindent{\bf Proof.}  By definition, $S^{(k)}\subseteq S^{(k-1)}\subseteq \{1,2\cdots,n\}$, for all $k\in \mathbb{N}^+$. Then, the sequence $\{|S^{(k)}|\}_{k=1}$ is decreasing and has an upper bound $n$.
 Assume that there does not exist a positive integer $k\leq n+1$ such that $S^{(k)}=S^{(k-1)}$. Then, $|S^{(k)}|<|S^{(k-1)}|$ for $k\leq n+1$ and $|S^{(1)}|<n$.
 Thereby, we have $|S^{(k)}|=0$ for all $k\geq n$, and therefore $|S^{(n)}|=|S^{(n+1)}|$. It leads a contradiction. The proof is complete.  \qed
\begin{Remark}
 In addition, ASDBR algorithm will be terminated when the maximum outer iteration $k_{\rm outer}$ is reached.
\end{Remark}

The following is an immediate consequence of Theorem \ref{th1}.
\begin{Corollary}\label{cor1}
Let $\{S^{(k)}\}$ be a support set estimate sequence generated from ASDBR algorithm. Then, before ASDBR algorithm reaches its termination condition, it can obtain an s-sparse estimate in at most $n-s$ iterations.
\end{Corollary}

\section{Experiment results}\label{se4}

In this section, the experiments are presented to verify the effectiveness of the proposed algorithm. ASDBR is also compared with popular methods, i.e., Lasso \cite{tib1996} and SBL \cite{pan2015sparse}.
To evaluate the performance, we use the root of normalised mean square error
(RNMSE) as a performance index:
\[
\mathrm{RNMSE}:=\|\widehat{\theta}-\theta_{\mathrm{true}}\|_2/\|\theta_{\mathrm{true}}\|_2,
\]
where $\widehat{\theta}$ is the estimate of true parameter vector $\theta_{\mathrm{true}}$. We define the signal-to-noise ratio (SNR) as
\[
\mathrm{SNR(dB):=20\log_{10}}\left(\|\Phi\theta_{\mathrm{true}}\|_2/\|w\|_2\right).
\]
In all experiments, the number of inner iteration $k_{\rm  inner}$ is set to 1000, the number of outer iteration $k_{\rm outer}$ is set to 10.
 For fair comparison, 100 independent experiments are conducted to obtain the average RNMSE and runtime.
\subsection{Problem specification}
In the first experiment, the dictionary matrix $\Phi$ is selected to be a $M\times N$ Gaussian random matrix, whose elements are generated from independently and identically distributed (i.i.d.) normal distribution with zero mean and variance 1. A sparse vector $\theta$ of length $N$ is generated such that $\|\theta\|_0=K$.
The support, i.e., the location of the $K$ nonzero elements is chosen randomly, and the values are chosen from different distributions:
\begin{itemize}
  \item[1)] Uniform $\pm 1$ random spikes (Sub-Gaussian).
  \item[2)] Zero mean and unit variance Gaussian.
\end{itemize}

 In these two cases, we apply the proposed algorithm to identify the sparse vector. Here, the regularization parameter $\lambda$ is manually optimized as 1.

 \subsection{Recovery performance}
\begin{figure}
  \centering
  \includegraphics[width=6cm]{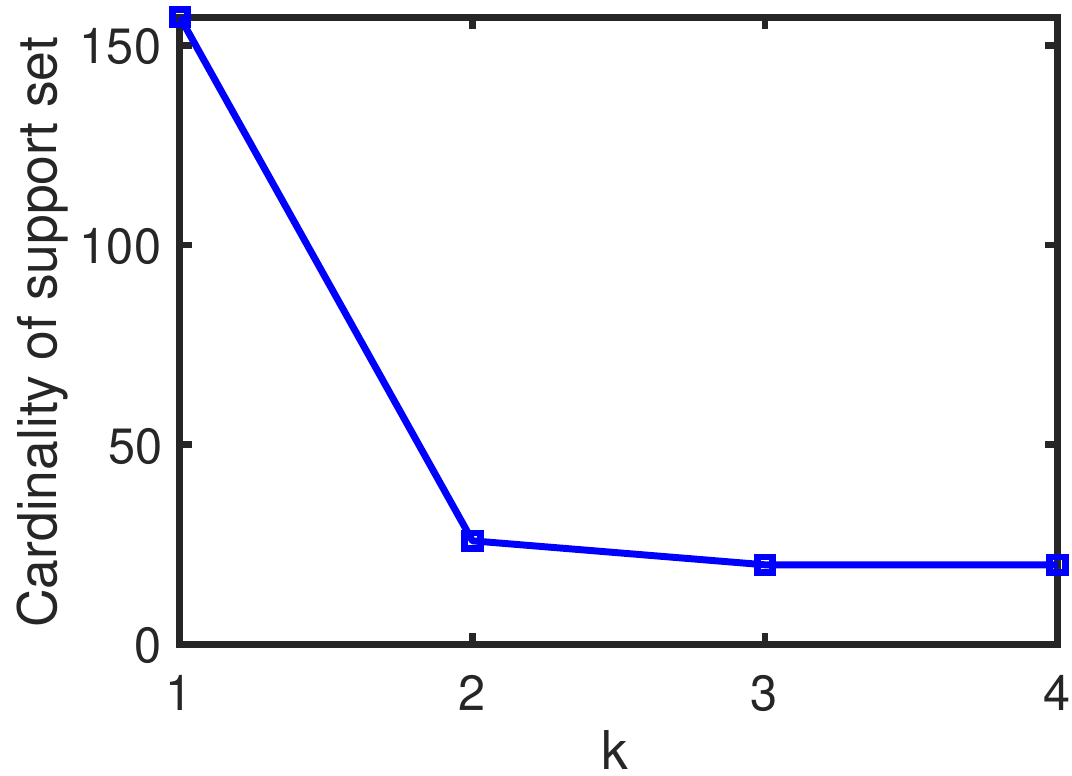}\\
  \caption{Cardinality of support set versus outer iteration steps.}\label{f1}
\end{figure}

To evaluate the performance of the proposed algorithm, we consider the case where $M=800$, $N=1600$, $K=20$ and $\mathrm{SNR}=15\mathrm{dB}$. Fig.~\ref{f1} shows the evolution of the cardinality of support set of the estimated vector along  outer iterations. From this figure, one see that ASDBR algorithm is terminated at \emph{4}th outer iteration, at which the cardinality of support set of the estimated vector and $\mathrm{RNMSE}$ are 20 and 0.0075, respectively. Thus, the proposed algorithm can recover the true sparse vector. Moreover, like the SBL algorithm, the main memory consumption of ASDBR algorithm depends on the size of matrix $\Phi^T\Phi$ (See \eqref{eq13}). Then, these algorithms can be compared by memory consumption. As shown in Fig.~\ref{f1}, for the proposed algorithm, at the first outer iteration, the size of the dictionary matrix $\Phi$ is greatly compressed from $800\times1600$ to $800\times 157$; at the second outer iteration, the size of the dictionary matrix $\Phi$ is compressed from $800\times157$ to $800\times 26$; at \emph{3}th outer iteration and \emph{4}th outer iteration, the corresponding dictionary matrix $\Phi$ has the same size $800\times 20$ and hence the iteration terminates. As a result, the memory consumption is reduced, which improves the computational speed.

 \subsection{ Comparison with other algorithms}
To demonstrate the advantage of the proposed algorithm, we compare it with Lasso and SBL in terms of $\mathrm{RNMSE}$ and runtime. Here we consider the following three cases.
\begin{itemize}
  \item[1)] $M/N=0.5$, $K=20$, $\mathrm{SNR}=10\mathrm{dB}$, and the value of $N$ ranges from 400 to 2000.
  \item[2)] $M=400$, $N=800$, $K=20$, and the value of $\mathrm{SNR}$ ranges from 0dB to 20dB.
  \item[3)] $M=400$, $N=800$, $\mathrm{SNR}=10\mathrm{dB}$, and the value of $K$ ranges from 10 to 60.
\end{itemize}

\begin{figure}
  \centering
  \includegraphics[width=7.5cm]{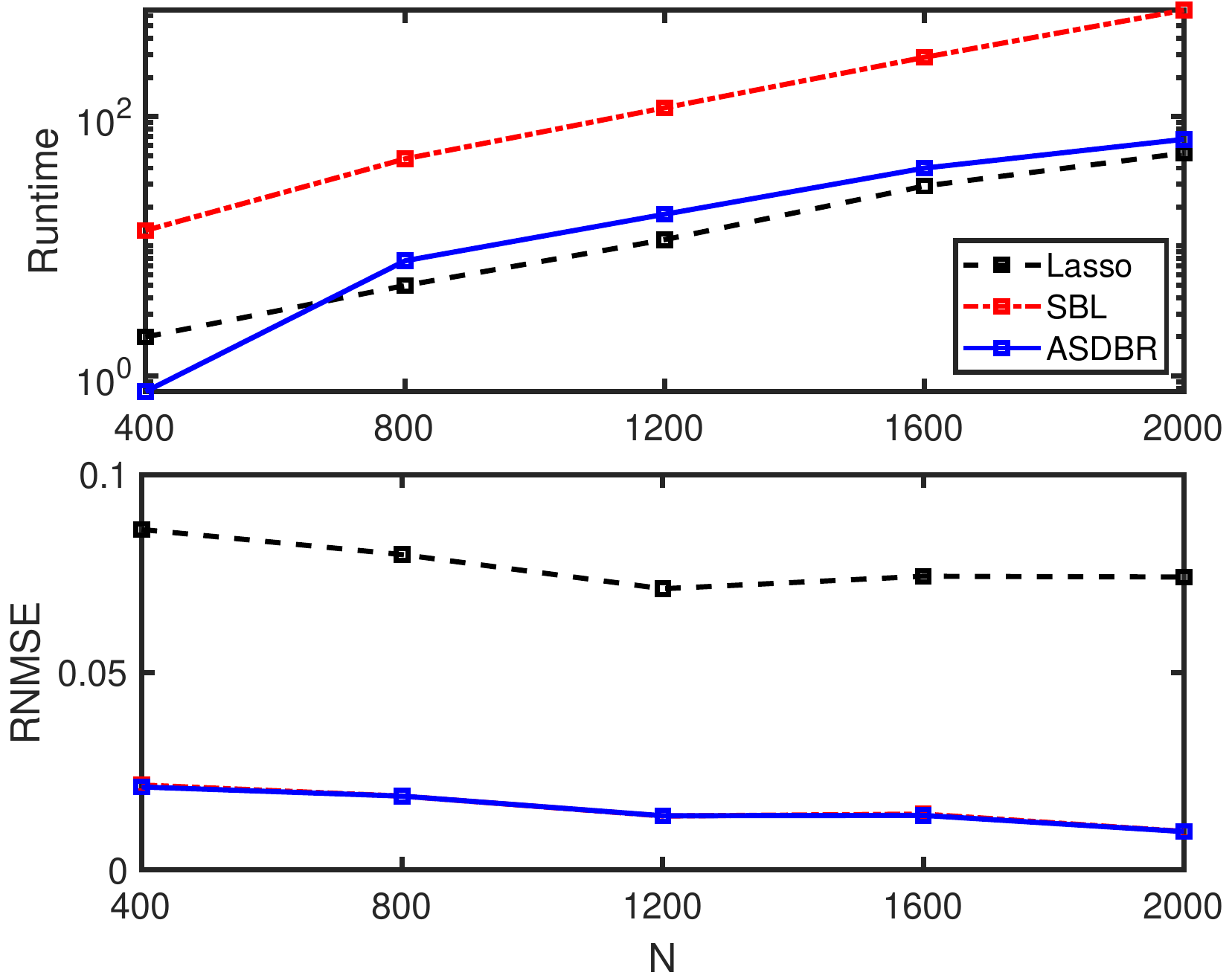}\\
  \caption{Runtime and RNMSE comparison for sub-Gaussian distributed nonzero coefficients versus problem dimension N.}\label{f3}
\end{figure}

\begin{figure}
  \centering
  \includegraphics[width=7.5cm]{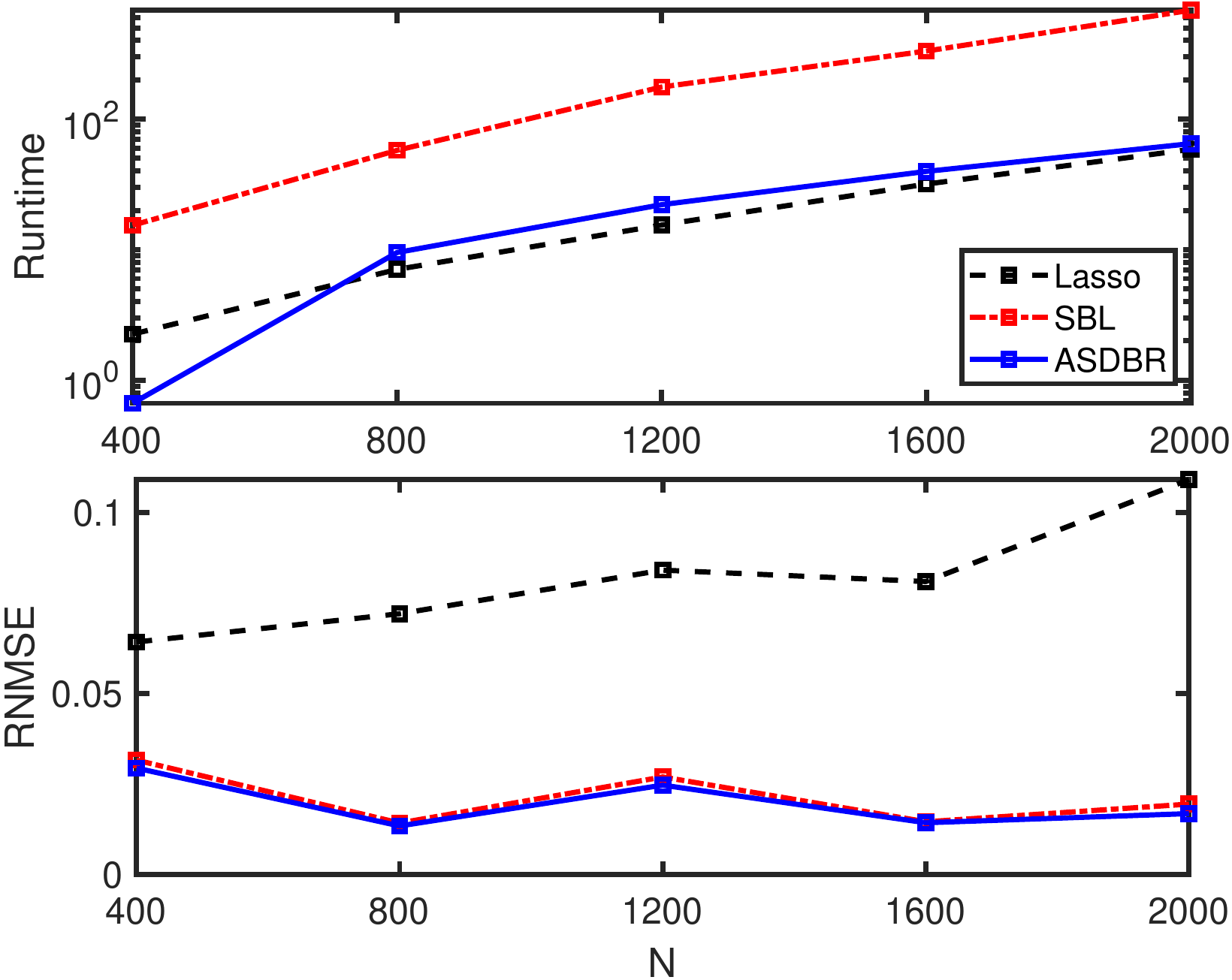}\\
  \caption{Runtime and RNMSE comparison for Gaussian distributed nonzero coefficients versus problem dimension N.}\label{f5}
\end{figure}

 In the first case, the evolutions of runtime and $\mathrm{RNMSE}$ with problem dimension $N$ are plotted in Fig.~\ref{f3} and Fig.~\ref{f5}. Therein, Fig.~\ref{f3} corresponds to the sub-Gaussian distributed nonzero coefficients, while Fig.~\ref{f5} corresponds to the Gaussian distributed nonzero coefficients. From these figures, one see that $\mathrm{RNMSE}$s of ASDBR and SBL are almost the same and less than that of Lasso, while ASDBR requires less runtime than SBL.

\begin{figure}
  \centering
  \includegraphics[width=7.5cm]{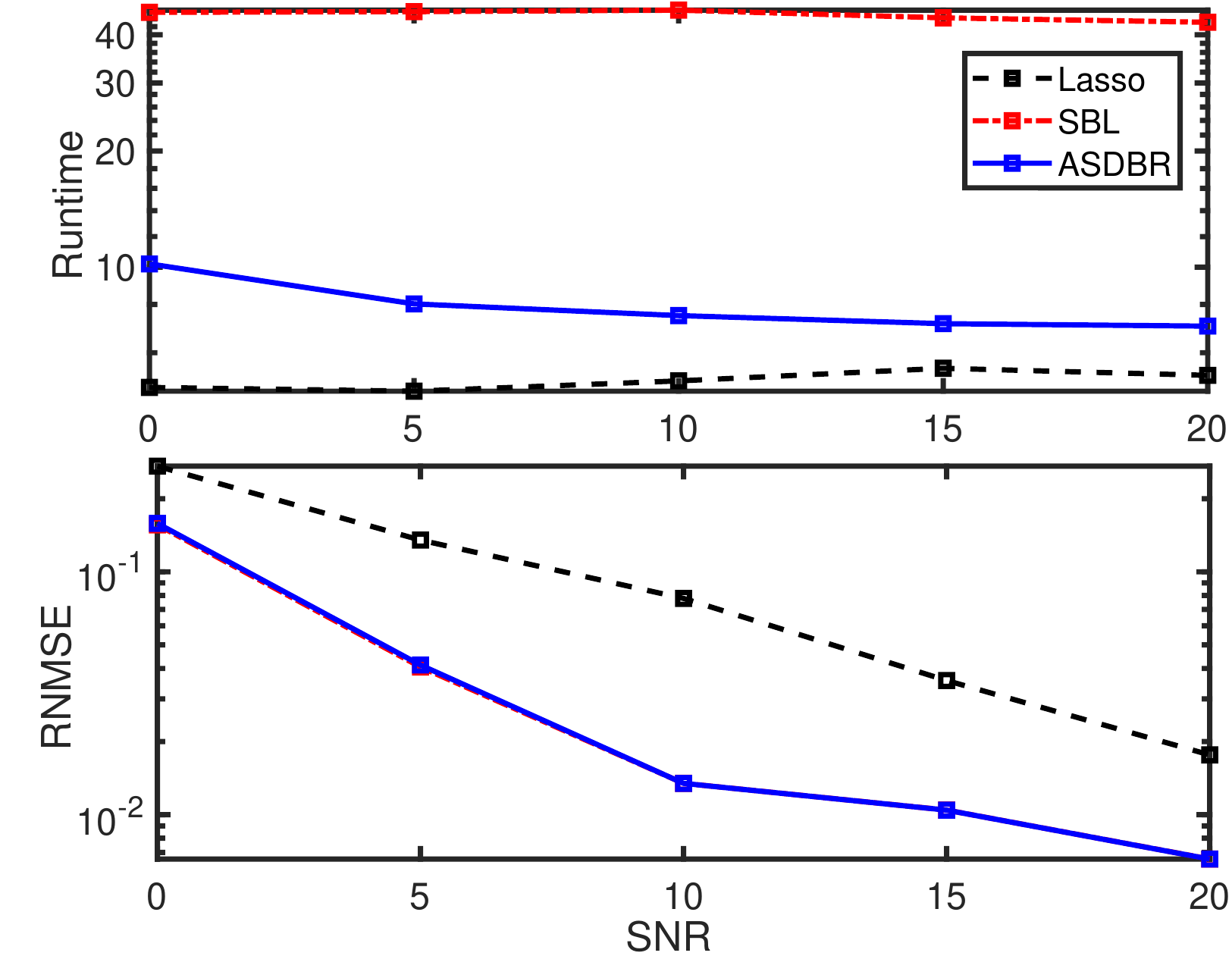}\\
  \caption{Runtime and RNMSE comparison for sub-Gaussian distributed nonzero coefficients versus SNR.}\label{f2}
\end{figure}

\begin{figure}
  \centering
  \includegraphics[width=7.5cm]{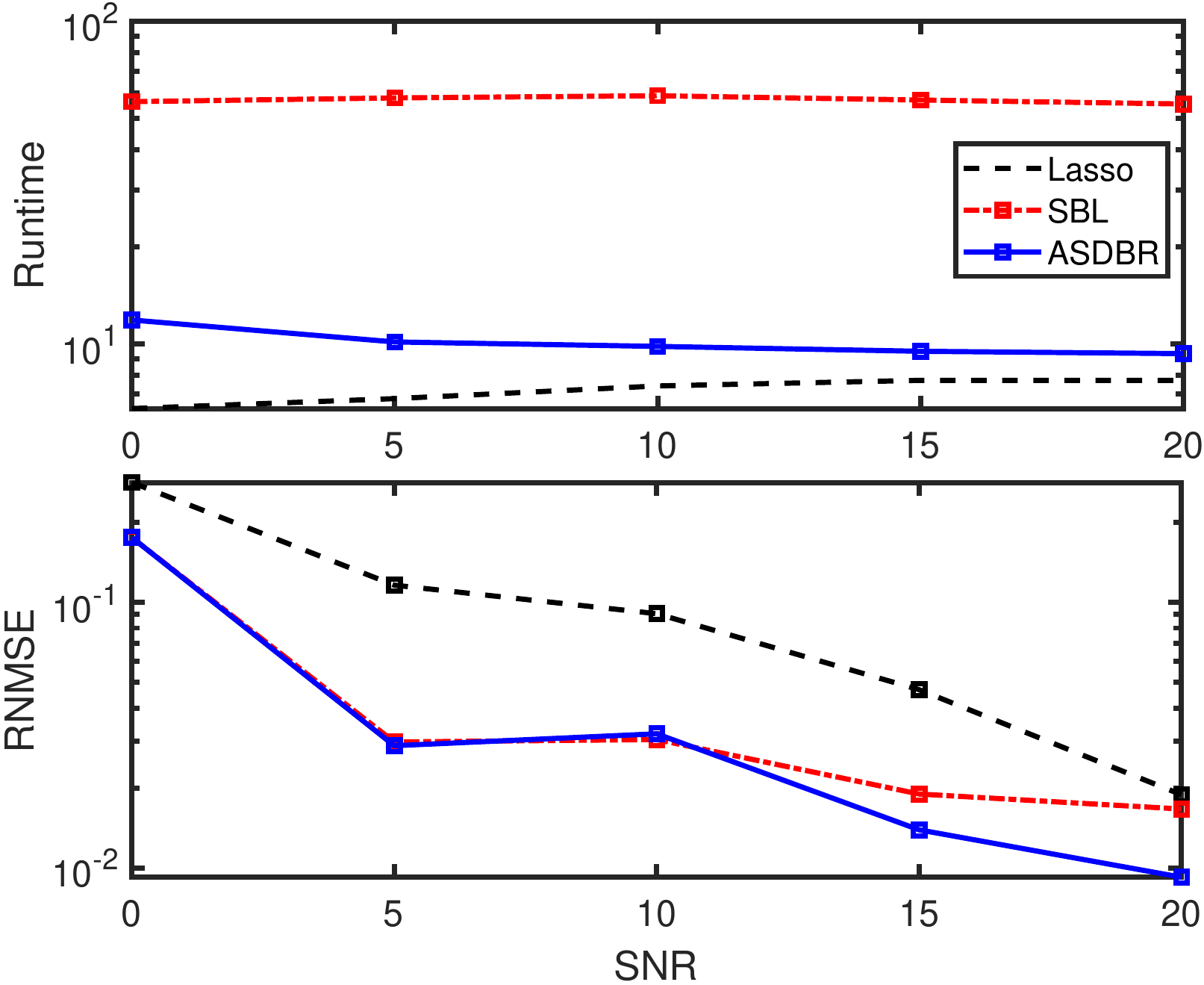}\\
  \caption{Runtime and RNMSE comparison for Gaussian distributed nonzero coefficients versus SNR.}\label{f4}
\end{figure}

 In the second case, the evolutions of runtime and $\mathrm{RNMSE}$ with $\mathrm{SNR}$ are plotted in Fig.~\ref{f2} and Fig.~\ref{f4}. From these figures, one see that ASDBR is close to or better than SBL, and both of them are better than Lasso, in terms of $\mathrm{RNMSE}$. Specially, for Gaussian distributed nonzero coefficients, when the SNR is relatively large, ASDBR is better than SBL in terms of $\mathrm{RNMSE}$. Moreover, observe that ASDBR requires less runtime than SBL.

\begin{figure}
  \centering
  \includegraphics[width=7.5cm]{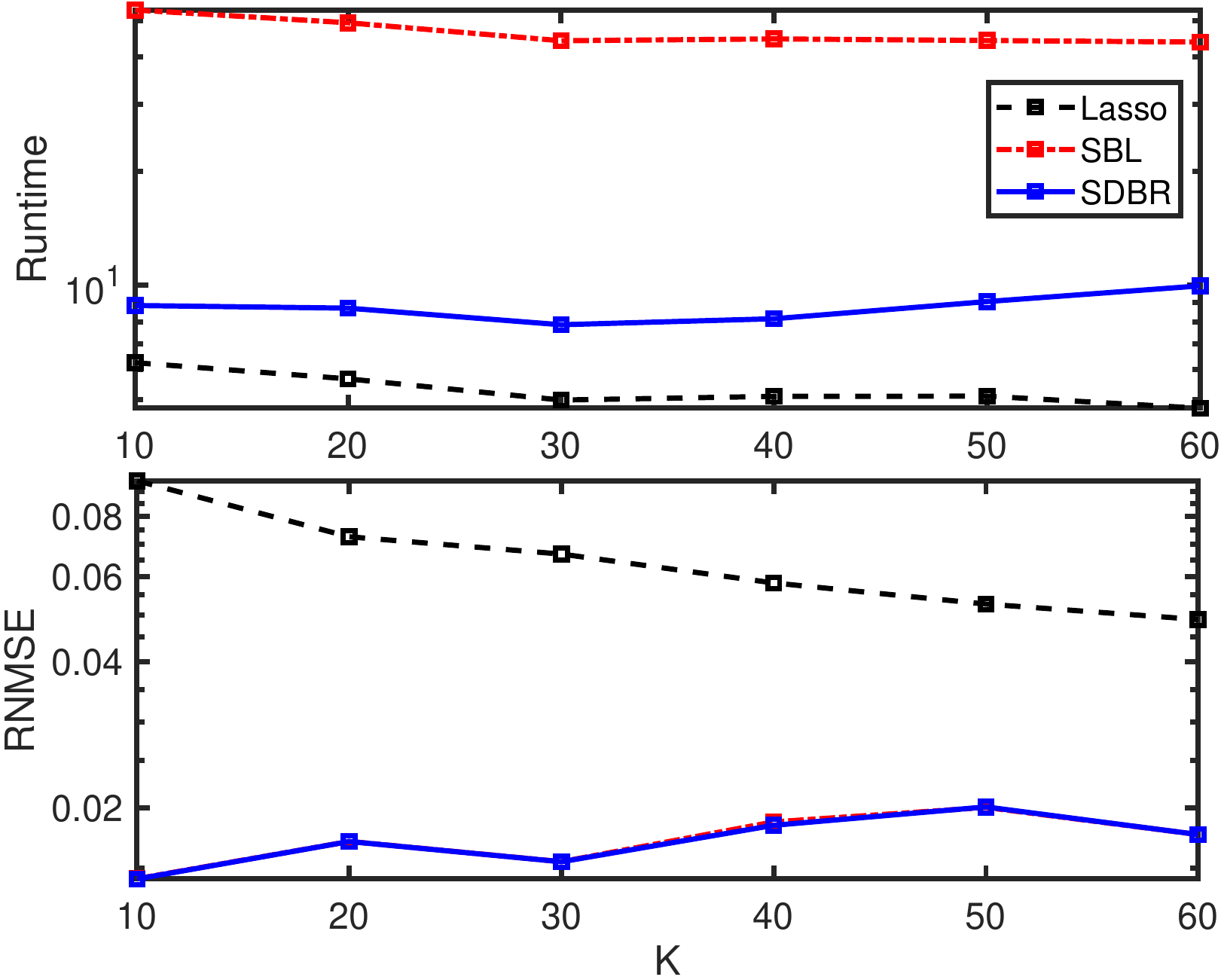}\\
  \caption{Runtime and RNMSE comparison for sub-Gaussian distributed nonzero coefficients versus sparsity K.}\label{f6}
\end{figure}

\begin{figure}
  \centering
  \includegraphics[width=7.5cm]{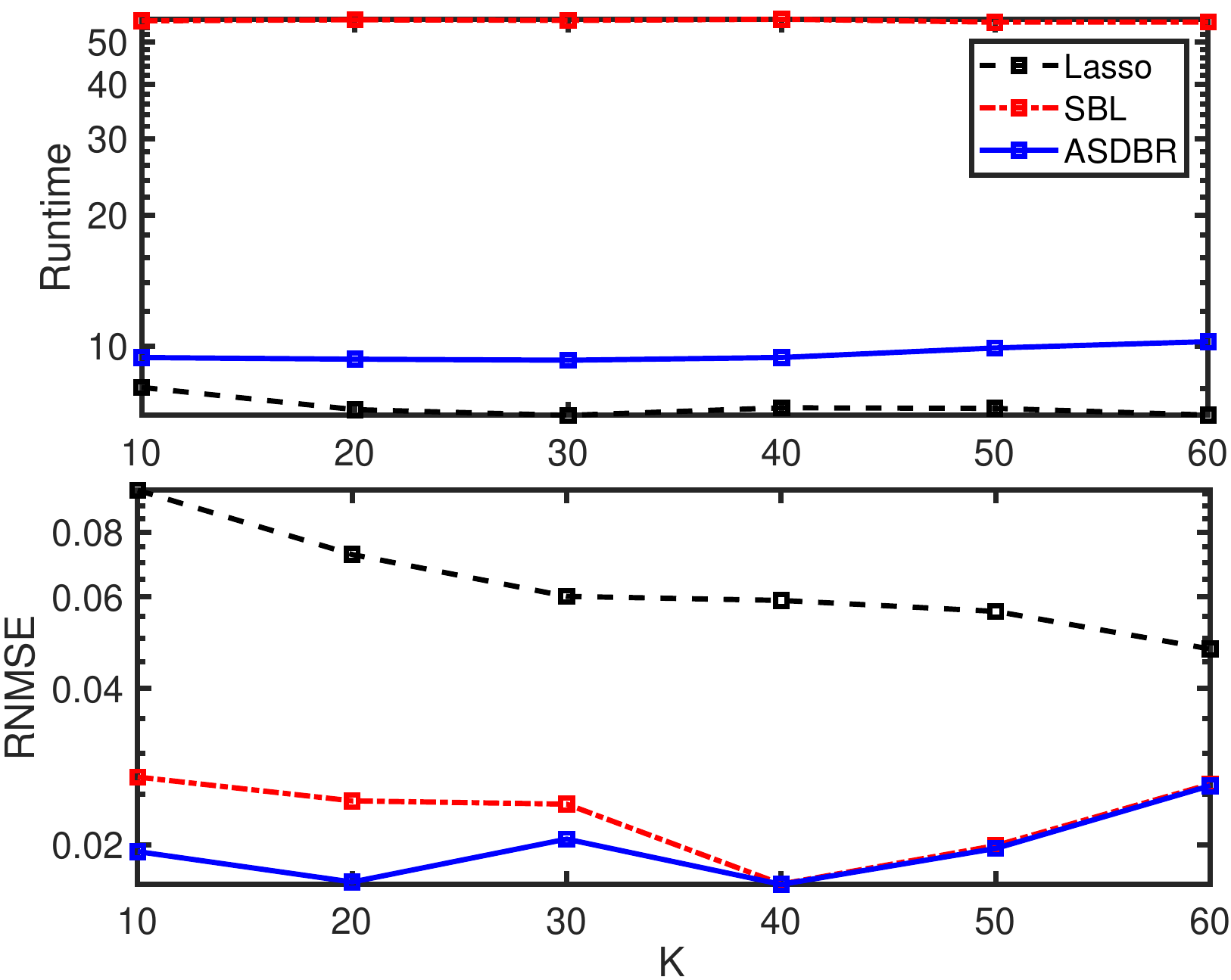}\\
  \caption{Runtime and RNMSE comparison for Gaussian distributed nonzero coefficients versus sparsity K.}\label{f7}
\end{figure}

 In the third case, the evolutions of runtime and $\mathrm{RNMSE}$ with sparsity $K$ are plotted in Fig.~\ref{f6} and Fig.~\ref{f7}. From these figures, one see that ASDBR is close to or better than SBL, and both of them are better than Lasso, in terms of $\mathrm{RNMSE}$. Specially, for Gaussian distributed nonzero coefficients, when $K$ is small, ASDBR is better than SBL in terms of $\mathrm{RNMSE}$. Note that, ASDBR requires less runtime than SBL.
\section{Conclusions}\label{se5}
In this paper, we developed a ASDBR method based on a Bayesian model for solving sparse recovery problem.
ASDBR uses a threshold strategy leading to a support estimate, which can increase computational speed with low memory consumption. The simulation results show the computational advantages of the proposed algorithm. 

\end{document}